\documentclass[doublecol]{epl2}% for 2 columns style without line numbers
\usepackage{amsmath,bm,epsfig}
\usepackage{epstopdf}
\usepackage{color}
\usepackage{ulem}
\newcommand{\B}[1]{{\bm{#1}}}%% Bold Roman & Greek Lower & Upper Case
\newcommand{\C}[1]{{\mathcal{#1}}}    %%   Calligrapfic Upper case

\title{The Static Lengthscale Characterizing the Glass Transition at Lower Temperatures}
\shorttitle{Static Lengthscale Characterizing the Glass Transition at Lower T} %Insert here a short version of the title if it exceeds 70 characters

\author{R. Guti\'errez \inst{1} \and S. Karmakar\inst{2} \and Y. G. Pollack\inst{1} \and I. Procaccia\inst{1}}

\institute{
  \inst{1} Dept. of Chemical Physics, The Weizmann Institute of Science -  Rehovot 76100, Israel\\
  \inst{2} TIFR Center for Interdisciplinary Science - Narsingi, Hyderabad, 500075, India.
}
\pacs{64.70.P-}{Glass Transitions}
\pacs{61.43.Fs}{Structure of Glasses}
\pacs{64.70.Q-}{Theory and Modeling of Glass Transitions}

\abstract{
The existence of a static lengthscale that grows in accordance with the
dramatic slowing down observed at the glass transition is a subject of
intense interest. Due to limitations on the relaxation times reachable by standard molecular dynamics
techniques (i.e. a range of about 4-5 orders of magnitude) it was until now impossible to demonstrate a significant enough increase in any proposed length scale. In this Letter we
explore the typical scale at unprecedented lower temperatures. A swap Monte Carlo approach
allows us to reach a lengthscale growth by more than 500\%. We conclude by discussing the relationship between the
observed lengthscale and various models of the relaxation time, proposing that the associated increase in relaxation time
approaches experimental values.}

\begin{document}

\maketitle

%%%%%%%%%%%%%%\section{Section title }

The most conspicuous aspect of the phenomenon referred to as
``the glass transition" is the immense increase in relaxation time of a supercooled
liquid upon a modest reduction of its temperature. It is not uncommon to
observe experimentally an increase of $10-14$ orders of magnitude in the
measured viscosity. The phenomenon of slowing down is correlated with
the dynamic heterogeneity  in super-cooled liquids which is seen in both
experiments and simulations~\cite{BookDH}. Dynamic heterogeneity
indicates the existence of a {\it dynamical} lengthscale which is
revealed by studying multi-point correlation
functions~\cite{95HH,99BDBG,02DFGP,00FP,05Chi4,PREchi4,09KDS,10KDS}; more particles move in a correlated way when the temperature of the supercooled liquid is reduced.
Although this is an interesting feature of supercooled
liquids, it is still not clear to what extent dynamic correlations are the
consequence or the primary origin of slow dynamics~\cite{09KDS}.

On the other
hand a major theoretical question that is still not fully answered is whether
the tremendous increase in relaxation time is accompanied by a corresponding
increase in a typical (static) lengthscale~\cite{11BB, 14KDS, 01Donth, 01DS} similarly to what is observed in critical phenomena. The identification of a
candidate lengthscale characterizing the glass
transition attracted considerable amount of attention, with two candidates
showing the most promise. The first method is based on the ``point-to-set" (PTS)
length~\cite{04BB,MM}. This length allows one to probe the spatial extent of
positional amorphous order; it was measured in several numerical simulations
~\cite{CGV,08BBCGV,SauTar10,BeKo_PS,HocRei12} and shown to grow mildly in the
(rather high) temperature regime investigated.

Another, superficially unrelated way to define a static scale (hereafter denoted as $\xi$) was announced
in Ref.~\cite{11KLP} and employed further in Ref.~\cite{12KP}. The relation between these two differently defined lengthscales was
studied carefully in Ref.~\cite{13BKP} with the conclusion that
they are in fact in full agreement, except that the point-to-set lengthscale
is accessible at higher temperatures whereas the length $\xi$ is
more accurately evaluated at lower temperatures.

The starting point
for the definition of the latter scale is the fact that the low frequency tail of the
density of states (DOS) of amorphous solids reflects the excess of plastic modes which do not exist in the density of
states of purely elastic solids~\cite{02TWLB,10Sok}. This excess of modes is
sometimes referred to as the `Boson Peak'~\cite{09IPRS}. Here and below a
`mode' refers to an eigenfunction of the underlying Hessian matrix
calculated at the local minimum of the potential energy function $U$ (the so called `inherent state'). The
Hessian matrix is defined as
${\cal H}_{ij}^{\alpha \beta} = \frac{\partial^2 U}{\partial x_i^{\alpha} \partial x_j^{\beta}}$ where
$x_{i}^{\alpha}$ denotes the $\alpha^{th}$ component of coordinate of
particle $i$. Recently~\cite{11HKLP} it was proposed that the eigenvalues
$\{\lambda_i\}_{i=1}^{dN}$ (with $d$ being the space dimension and $N$ the number of particles) appear
in two distinct families in generic amorphous solids, one corresponding
to eigenvalues of the Hessian matrix that are only weakly sensitive to
external strains; the other group consists of eigenvalues that go to zero at
certain values of the external strain, thus leading to a plastic failure.
The first group of modes is decently described by the Debye model of an
elastic body, but this is not the case for the second group corresponding
to the density of plastic modes. A simple model for the full density of
states was suggested~\cite{11HKLP,10KLP} as a sum of the Debye and the plastic
modes in the form
	\begin{equation}
		\label{Poflam}
		P(\lambda) \simeq A\left( \frac{\lambda}{ \lambda_D } \right)^{\frac{d-2}{2}}
		+ B(T) f_{\rm pl}\left( \frac{\lambda}{ \lambda_D } \right) \ .
	\end{equation}
where the pre-factor $B(T)$ depends on the temperature,
$\lambda_D \simeq \mu \rho^{2/d - 1}$ is the Debye
cutoff frequency and $\mu$ is the shear modulus. Particular models for the
function $f_{\rm pl}\left(\frac{\lambda}{ \lambda_D }\right)$ were presented
in~\cite{10KLP}. For our purposes here it is only
important to understand that this function is a partial characterization of
the degree of disorder which decreases upon approaching the glass transition.

The idea to determine the static typical scale
is that the {\it minimal} eigenvalue $\lambda_{\rm min}$ observed in a system
of $N$ particles will be determined by either the first {\it or}
the second term in Eq.~\eqref{Poflam}. For a system large enough, local
disorder will be irrelevant in determining $\lambda_{\rm min}$, and it will
be determined by the Debye contribution. For small systems the opposite is
true. Thus there exists a system size where a cross-over occurs. This
cross-over is interpreted in terms of a typical lengthscale $\xi$ separating
correlated disorder from asymptotic elasticity.

The calculation of the length $\xi$ at various T requires equilibrating the supercooled
liquid. For this reason,
until now one could only compute $\xi$ over a range
that corresponded to a change in the $\alpha$ relaxation time $\tau$ of up to $4-5$ orders of magnitude.
In this Letter we explore a ``swap Monte Carlo"~\cite{01GP}
technique to reach a measurement of $\xi$ over a range that corresponds
to a change in $\tau$ of between 12 and 16 orders of magnitude (depending of the form of the extrapolation or the functional dependence of $\tau$ on $\xi$ considered). The system under study is a ternary mixture of $N$ point particles made of $15\%$ particles A, $30\%$ particles B and $55\%$ particles C, which interact via a pure repulsive soft sphere potential similar to that in Ref.~\cite{01GP}:
\begin{equation}
	U({\bf r_{ij}}) = \left\{
	\begin{array}{ll}
		\epsilon  \left (\frac{\sigma_{ij}}{r_{ij}}\right)^{12}  + \sum_{l=0}^2
		c_{2l} \left(\frac{r_{ij}}{\sigma_{ij}}\right)^{2l} & \quad
		\frac{r_{ij}}{\sigma_{ij}} \leq x_c, \\ \\
		\quad 0 & \quad \frac{r_{ij}}{\sigma_{ij}} > x_c.
	\end{array}
	\right.
\end{equation}
Here, $\sigma_{ij} = \sigma_i + \sigma_j$, where $\sigma_i$ is determined by the particle type,
and  $r_{ij}\equiv |\B r_i-\B r_j|$. The energy scale $\epsilon$ as well as the particle mass $m$ are taken as unity. Time units are given by $\sqrt{m\, l_0^2/\epsilon}$, where $l_0$ gives the unit of length. The coefficients $c_{2l}$ are chosen in such a way that the potential and
its first and second derivatives vanish at the cutoff $x_c = 1.25$.
We choose $\sigma_A$,
$\sigma_B$ and $\sigma_C$ such that $\sigma_A/\sigma_B = \sigma_B/\sigma_C =
1.25$, with $\sigma_C=0.424741977632123\, l_0$. This guarantees that the
average value of the diameter of the dominating $\left(\frac{\sigma_{ij}}{r_{ij}}\right)^{12}$
soft sphere part of the potential  is unity.
The simulations are performed in the $NVT$ ensemble at
a density $\rho=1.1$ with periodic boundaries in 3D.
The use of three components and the choice of the
compositions and scales are designed to avoid the crystallization
observed in the original binary system which was reported elsewhere
~\cite{04BR,07FMV}. This system does not show a tendency to
crystallize at any of the temperatures examined for the time spans considered, as evidenced in the profile of the radial distribution functions and by the absence of abrupt transitions in the potential energy time series.

\begin{figure}
	\onefigure[scale=0.22]{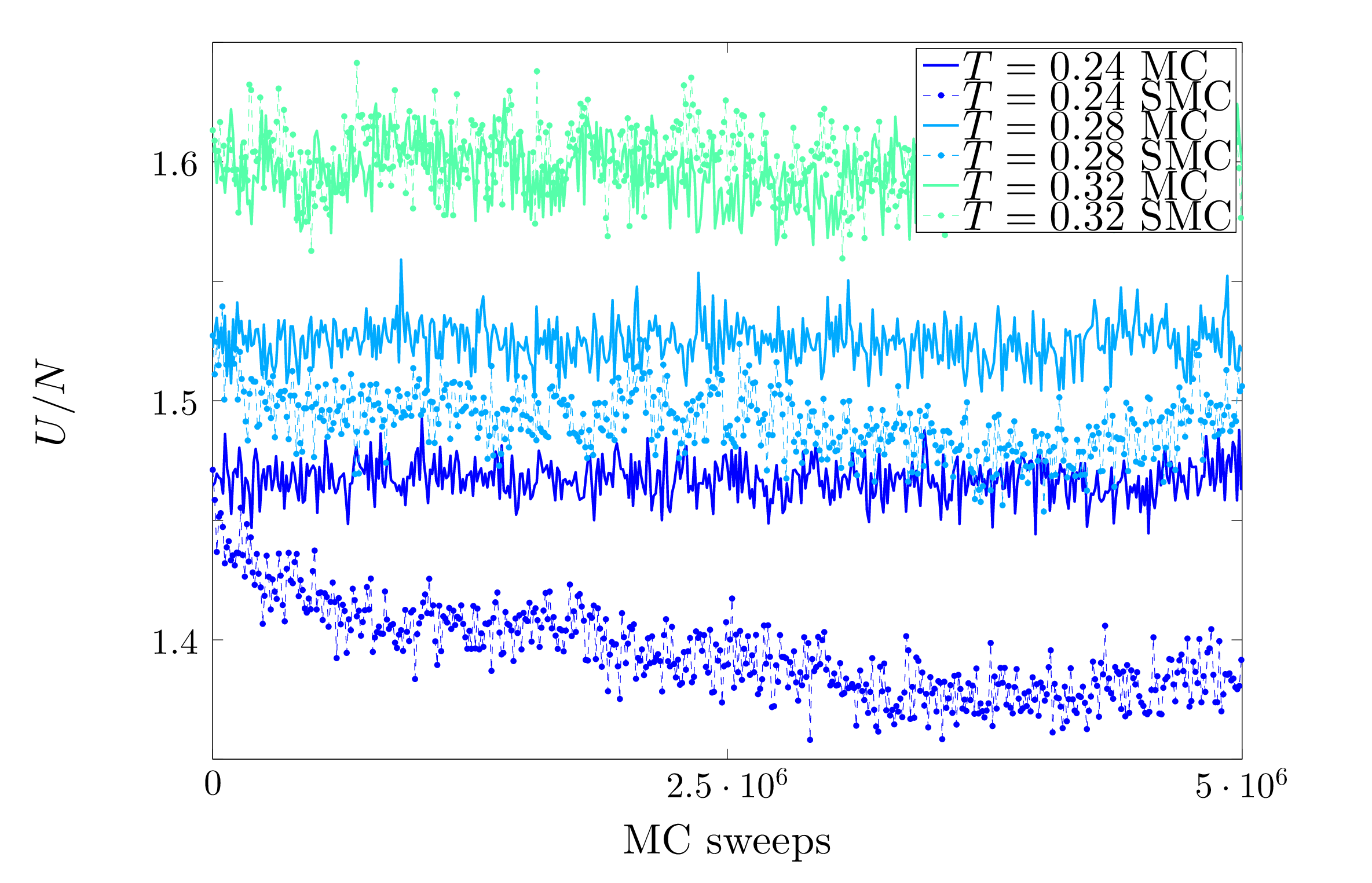}
	\caption{Time series of the potential energy per particle in a system of $N=1474$ particles for $T=0.32$ (green), $T=0.28$ (light blue) and $T=0.24$ (dark blue). Results obtained with regular Monte Carlo (MC) (continuous line) and swap Monte Carlo (SMC) (dash-dotted lines) simulations. Similar results were obtained for larger system sizes (not shown here) }
	\label{efficiency}
\end{figure}

The Swap Monte Carlo method combines standard Monte Carlo moves with particle swaps. Both types of moves are
accepted or rejected according to the standard Metropolis algorithm (detailed balance holds for both types of moves under this criterion). While particle swaps are accepted at lower rates as the
temperature is decreased, they nonetheless greatly accelerate equilibration by providing a way for
the particles to break away from the cages in which they would be otherwise
stuck, at least as long as the acceptance rate is of the order of $10^{-4}$ or
even $10^{-5}$. An illustrating comparison between the regular and swap Monte Carlo methods is provided in Fig.~\ref{efficiency}, where it is shown that Swap Monte Carlo is able to relax the system at such low temperatures for which regular Monte Carlo shows no perceptible departure from the initial configuration after millions of sweeps. These results are in qualitative agreement with those observed for a binary mixture of soft particles in Ref.~\cite{01GP}.
	
\begin{figure}
	\onefigure[scale=0.18]{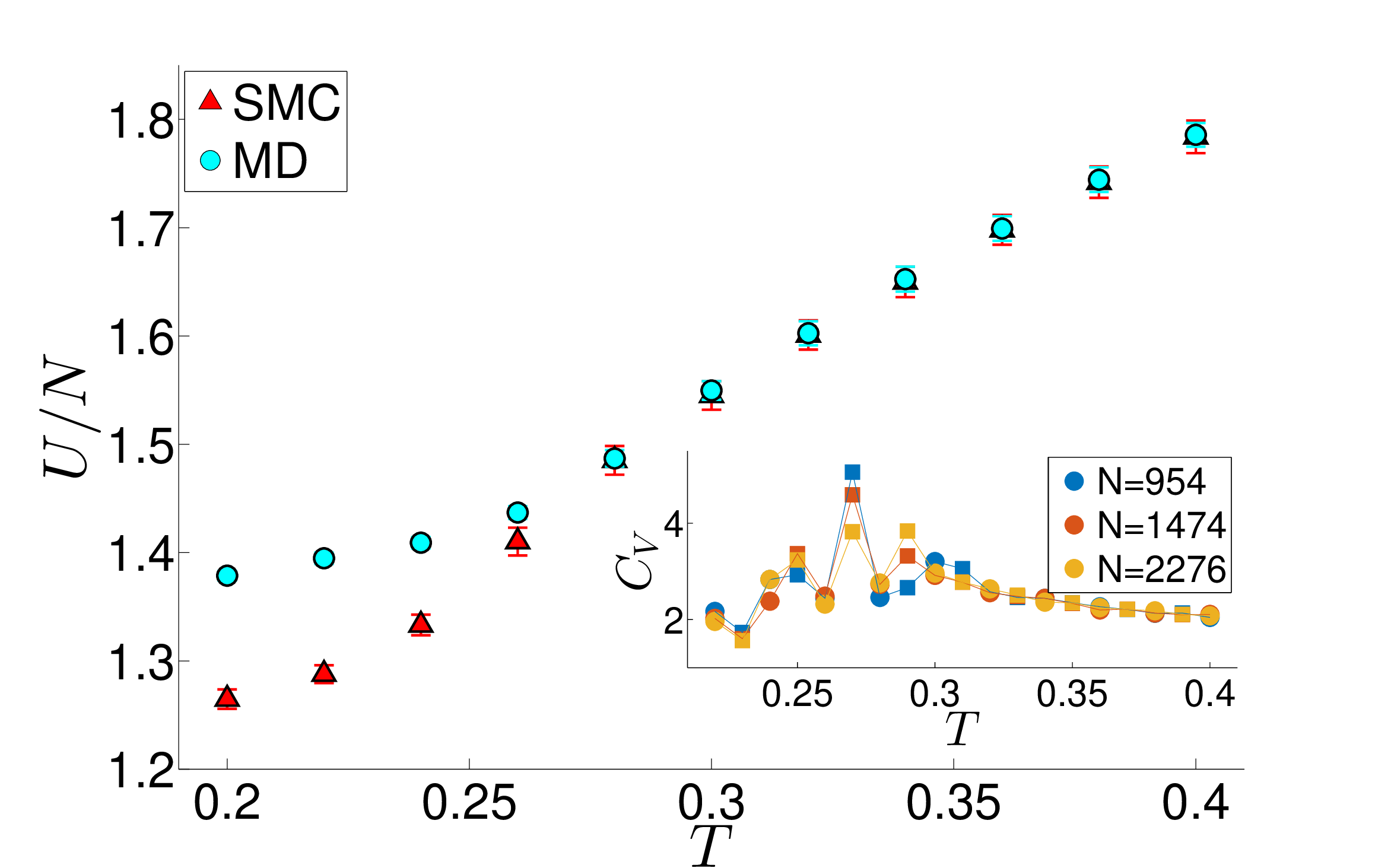}
	\hspace{-0.6cm}
	\onefigure[scale=0.24]{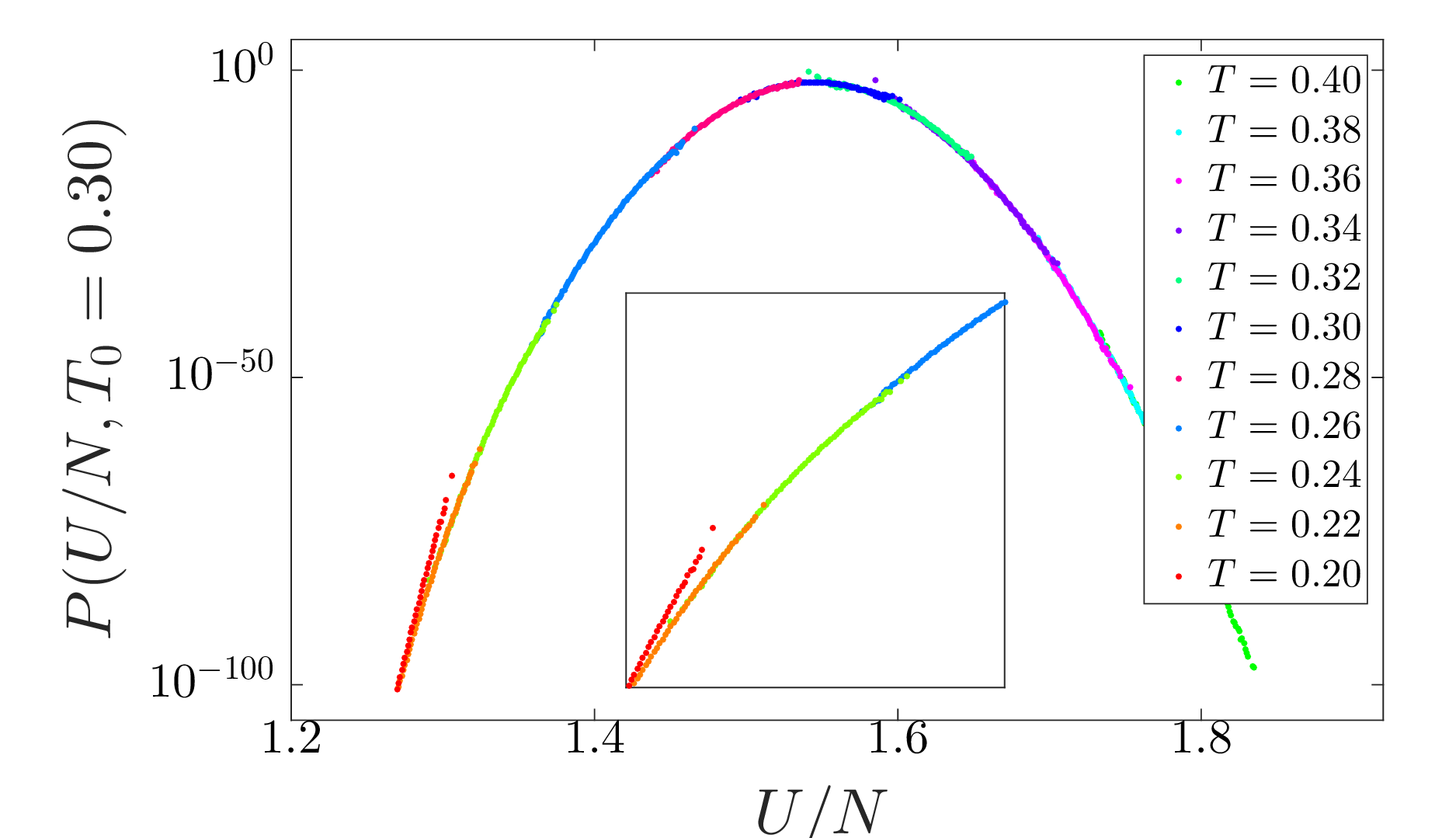}
	\caption{Upper panel: The potential energy per
		particle of the supercooled liquid as a function of temperature.
		Circles: standard molecular dynamics (MD). Departure
		from the equilibrium curve is observed at $T\approx 0.26$. Triangles:
		Swap Monte Carlo(SMC) with equilibration down to lower temperatures. Inset: specific heat $C_V$ computed from the energy standard deviation (circles) and from the energy derivative (squares). Lower panel: Collapse of the potential energy distribution $P(U,T)$ for $T=0.20,\ldots,0.40$ on the distribution $P(U,T_0)$ corresponding to $T_0=0.30$ for $N=1474$. Equilibration is achieved for $T\geq0.22$. The inset shows the transition from the out-of-equilibrium $T=0.20$ distribution to the equilibrated $T=0.22$ distribution, and from this to $T=0.24$.}
	\label{eq_check}
\end{figure}

In order to study the typical lengthscale characterizing the glass transition, we need to define the range of temperatures at which we are able to equilibrate our system. For this task we consider molecular dynamics and Swap Monte Carlo. Regular Monte Carlo is not considered in the following for its relative inefficiency with respect to Swap Monte Carlo, as illustrated above. The upper panel of Fig.~\ref{eq_check} shows the low T limit for which equilibrium can be attained via molecular dynamics, as evidenced by the deviation of the low T potential energy from the extrapolated supercooled liquid curve. The initial configurations are prepared by equilibrating a fluid with molecular
dynamics at $T=0.5$ for $1000$ time units, and then cooling it down to the target
temperature with a rate of $3.33\times 10^{-4}$, a procedure that is also followed to generate the initial conditions for the Swap Monte Carlo simulations. The potential energy was measured across a time window of $50000$ time units after an equilibration time of the order of ${10^7}$ time units. Despite the enormous computational
efforts this required (it took several weeks of CPU time), it becomes impossible to equilibrate the system for $T \leq 0.26$, as the conspicuous departure from the supercooled liquid curve
indicates. Additionally, during the same time window in which the potential energy was measured, we computed the self-intermediate scattering function,
\begin{equation}
	\label{fsqt}
	F_s(q,t) = \frac{1}{N}\sum_{i=1}^{N}\exp{\left[\B q\cdot\left(\B r_i(t) - \B r_i(0)\right)\right]},
\end{equation}
where $\B r_{i}(t)$ is the position of the $i^{th}$ particle at time $t$
and $\B q$ is a wave vector. We choose the length of $\B q$ to correspond to the first peak of the structure factor, and we average across 50 orientations (as orientation should not be of relevance in an isotropic system). Every time $F_s(q,t)$ decays to $1/e$, a `relaxation instance' is said to occur, and $F_s(q,t)$ is restarted to $1$ (the current configuration is taken as $t=0$ reference configuration for the next decay). The relaxation time  $\tau$ reported below in Figs.~\ref{tauScaling} and \ref{fits} is computed as the average duration of a relaxation instance.

As for the Swap Monte Carlo results, the arrival to equilibrium was assessed by testing the attainment of
a canonical energy distribution as described below (cf. Fig.~\ref{eq_check} lower panel). This was done for temperatures as low as ${T = 0.22}$, the equilibration time for even lower temperatures being unreasonably long. Additionally we include ${T=0.20}$ in the equilibration analysis to illustrate the failure of equilibration. The resulting potential energies are shown in Fig.~\ref{eq_check} upper panel and can be compared to those of molecular dynamics.

For the study of the static
lengthscale below it is crucial to ascertain that our system is indeed equilibrated for $T\ge 0.22$.
For this purpose, we follow Ref.~\cite{00YK} and collapse the probability distribution of the potential energy $P(U,T)$ for different temperatures on the probability distribution of a reference temperature ${T_0=0.30}$.  The
energy distribution is rescaled according to
\begin{equation}
	\label{Updf_ansatz}
	P_{T_0}(U,T) = \frac{P(U,T) \exp{\left[(1/T - 1/T_0)U\right]}}{\int dU'
		P(U',T) \exp{\left[(1/T - 1/T_0)U'\right] }},
\end{equation}
which for equilibrated systems rescales the distribution to a canonical
one at a reference temperature $T_0$. The rescaled $P_{T_0}(U,T)$ for
$T=\left\{0.20,0.22,0.24,\ldots,0.40\right\}$ with $T_0=0.30$ are
shown in the lower panel of Fig.~\ref{eq_check}. The excellent collapse confirms that the
Swap Monte Carlo technique allows us to equilibrate the system at $T \ge 0.22$. In order to strengthen
the evidence for attaining equilibrium for all $T\ge 0.22$ we have checked that our system reached
detail balance for all these temperatures. We determined the Gaussian pdf for energy fluctuations
at each temperature $T$, and computed the variance of this pdf. Detailed balance is guaranteed when
the variance is $\sigma^2=k T^2 C_V$ where $C_V$ is the specific heat. Computing $C_V$ from $\partial \langle U \rangle/ \partial T$ we ascertained agreement to high precision, cf. inset in Fig.~\ref{eq_check} upper panel.
For lower temperatures even the present technique fails to equilibrate the system within reasonable
time (see inset of  Fig.~\ref{eq_check}, lower panel).
While a decrease in temperature from $T=0.28$
to $T=0.22$ may not seem impressive to the uninitiated eye, we will show now
that it has a major effect on the typical length.

The Swap Monte Carlo method is used to obtain a representative sampling of configuration space for a given temperature $T$ in the range for which equilibration is attainable. The decay of the intermediate scattering function allows us to focus on sufficiently uncorrelated configurations (i.e., configurations separated by one relaxation instance). These configurations are then quenched to zero temperature using conjugate gradient minimization~\cite{07PTVF}, landing them on an inherent state. We accumulated $1500$ inherent states for each temperature with
systems of sizes ranging from $N=954$ to $N=20000$. The minimal eigenvalue of the Hessian matrix is calculated at each of these inherent states using the Lanczos algorithm~\cite{96GL}.

Note that in systems of finite size both  $\lambda_{\rm min}$
and $\lambda_D$ are fluctuating from one amorphous realization to the other,
and we therefore consider their mean over realizations, denoted as
$\langle \lambda_{\rm min}\rangle $ and  $\langle \lambda_D\rangle $.

\begin{figure}
	\onefigure[scale = 0.24]{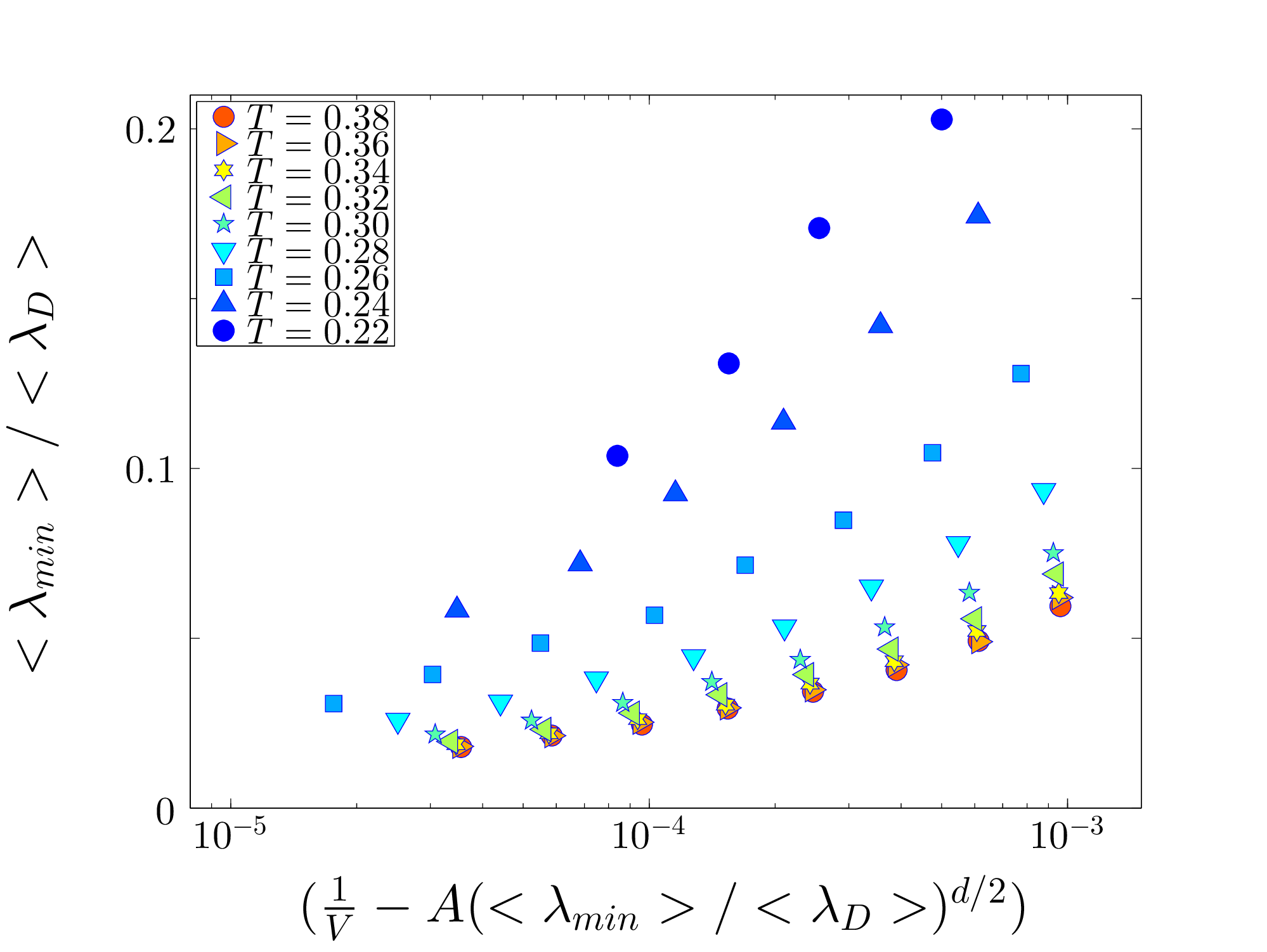}
	\onefigure[scale = 0.24]{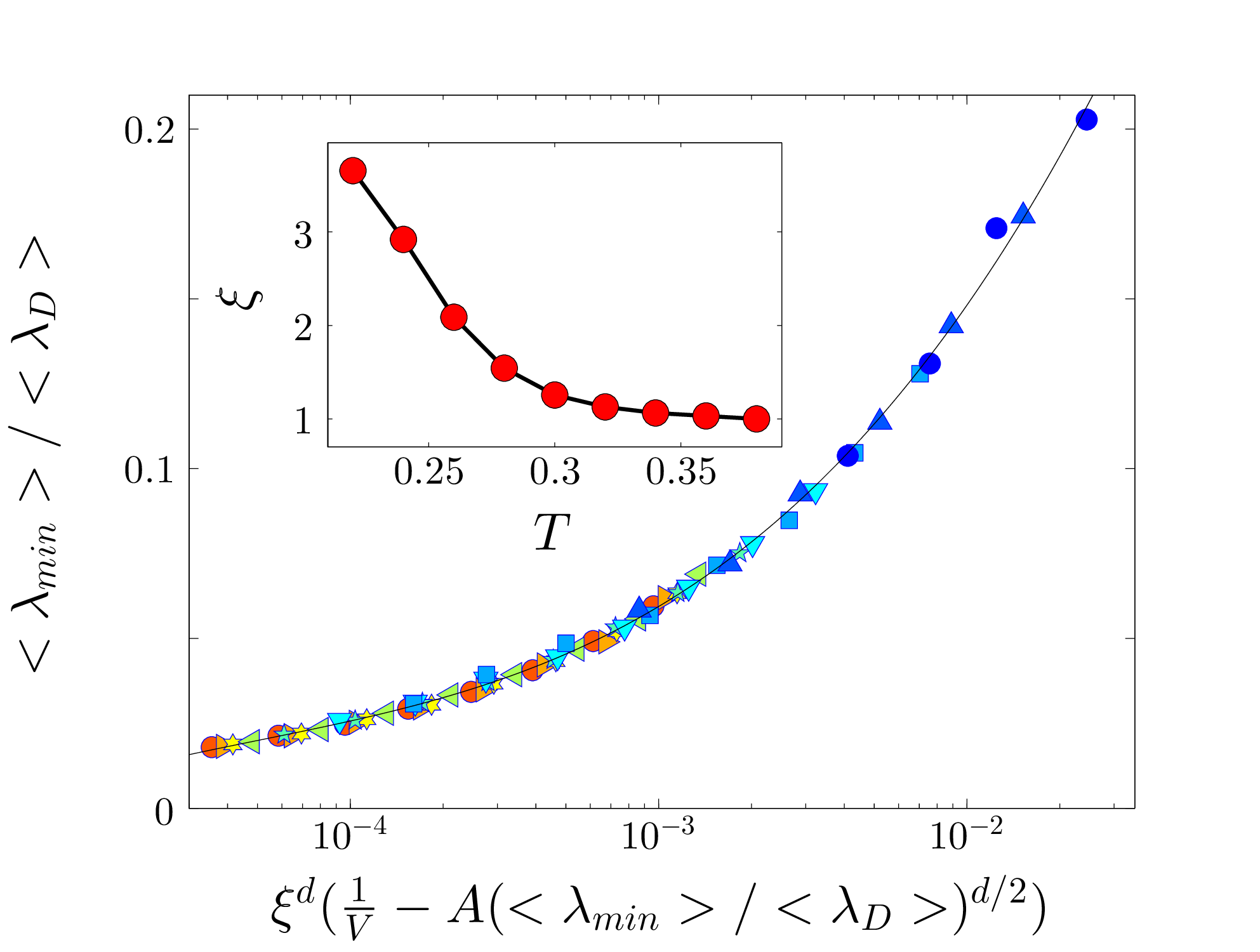}
	\caption{Upper Panel: The  function $\C F$ of eq.~\ref{ansatz} without
		rescaling for all the system sizes and all the temperatures. Lower Panel:
		Data collapse of the same data set by choosing the lengthscale $\xi(T)$. Inset: The temperature dependence of the extracted lengthscale. Due to the indeterminacy of an absolute scale we choose the lengthscale (for this figure only) to be unity at $T = 0.38$ . }
	\label{scaling}
\end{figure}
	
To proceed we integrate the density of states (Eq.~\eqref{Poflam}) from zero
to the average of the first eigenvalue. This integral picks up (on the
average) one out of $dN$ eigenvalues.
Because we integrate up to the average $\langle \lambda_{\rm min}
\rangle/\langle \lambda_{\rm D} \rangle$ we do not get
exactly 1 but ${\cal O} (1)$ in a finite system. (In an infinite system the
exact $1$ is recaptured). Explicitly,
%\begin{equation}
$Nd \int_{0}^{\frac{\langle \lambda_{\rm min} \rangle}{\langle \lambda_{\rm D}
		\rangle}} P\left(x\right)dx = {\cal O} (1)$.
%\end{equation}
Introducing Eq.~\ref{Poflam} into the integral
and doing some simple algebra one can show

\begin{equation}
{\cal G}\left(\frac{\langle\lambda_{\rm min}\rangle}{\langle\lambda_{\rm
		D}\rangle}\right) =
\left[ \frac{1}{\rho \tilde B(T)}\left(\frac{1}{V} - \tilde{A}
\left(\frac{\langle\lambda_{\rm min}\rangle}
{\langle\lambda_{\rm D}\rangle}\right)^{d/2} \right) \right] \ .
\end{equation}
where
${\cal G}\left( \frac{\langle\lambda_{\rm min}\rangle}{\langle\lambda_{\rm
		D}\rangle}\right) \equiv
\int_{0}^{\frac{\langle\lambda_{\rm min}\rangle}{\langle\lambda_{\rm
			D}\rangle}} f_{pl}(x)dx$ and $\tilde A$ and $\tilde B$ are the same as $A$ and $B$ up to absorbed $T$-independent constants.
Changing slightly this equation, writing  $\xi^d(T) \equiv \frac{1}{\rho \tilde
	B(T)}$, one obtains
the following scaling function
\begin{equation}
\label{ansatz}
\frac{\langle\lambda_{\rm min}\rangle}{\langle\lambda_{\rm D}\rangle} = {\C	F}\left[ \xi^d(T)\left(\frac{1}{V} -\tilde A  \left(\frac{\langle\lambda_{\rm min}\rangle}{\langle\lambda_{\rm	D}\rangle}\right)^{d/2} \right) \right].
\end{equation}
where $\C F\equiv \C G^{-1}$. Since the function $\C G$ is monotonically increasing, its inverse is well defined. The typical scale $\xi(T)$ is calculated by
demanding that all the data obtained for different system sizes and
temperatures should collapse into a master curve just by appropriately
choosing the $\xi(T)$.

The raw results for the present method, before rescaling, are shown in the
upper panel of Fig.~\ref{scaling}. The same results after rescaling by
$\xi(T)$ are shown in the lower panel of the same figure, and the
resulting typical scale as a function of $T$ is shown in
in the inset.
It is evident that the addition of the three (lowest temperature) points to the typical scale $\xi(T)$,
results in an increase of $\xi(T)$ by a substantially larger factor	than that reached by direct molecular dynamics simulations~\cite{12KP}.

\begin{figure}
	\onefigure[scale = 0.32]{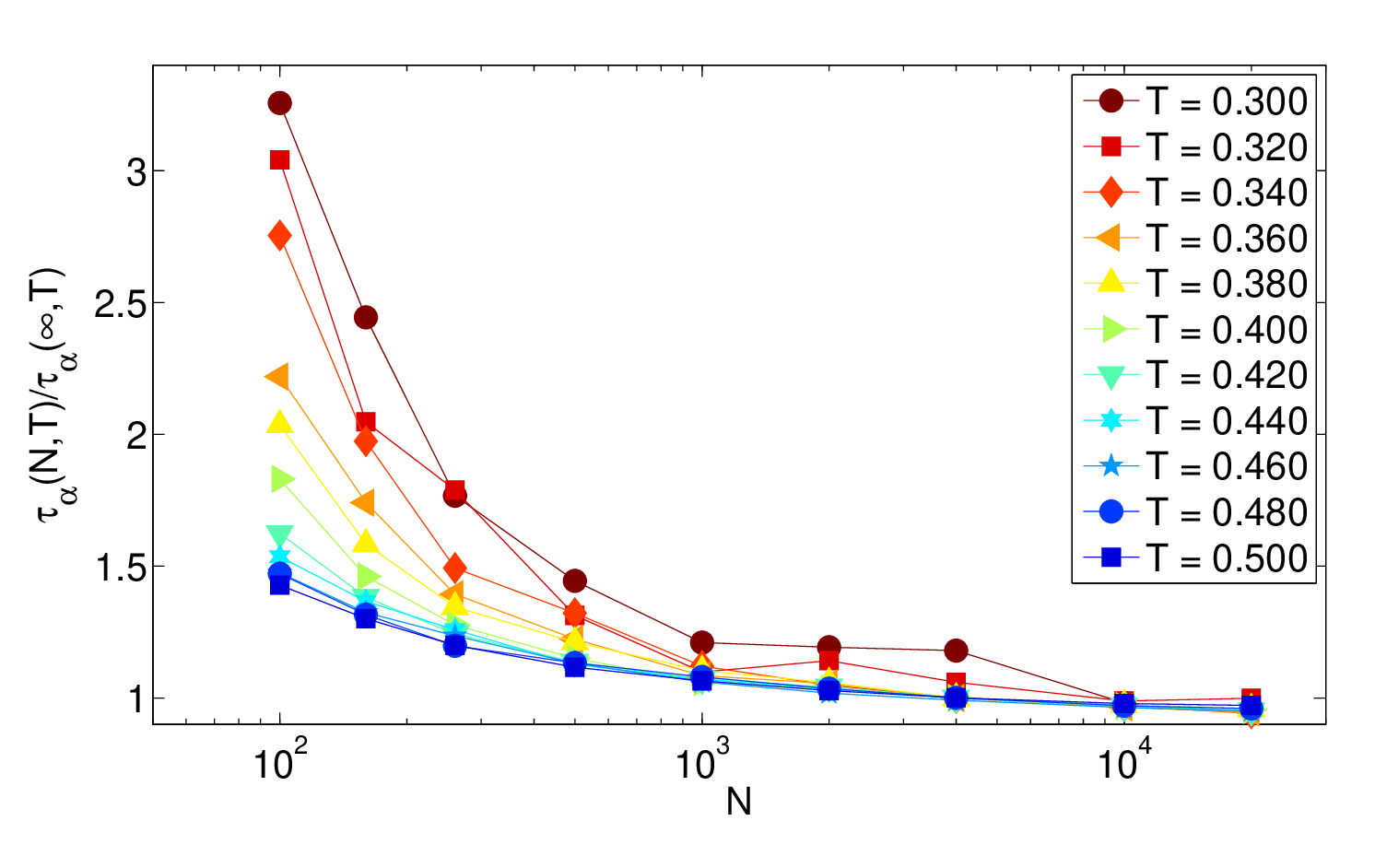}
	\onefigure[scale = 0.32]{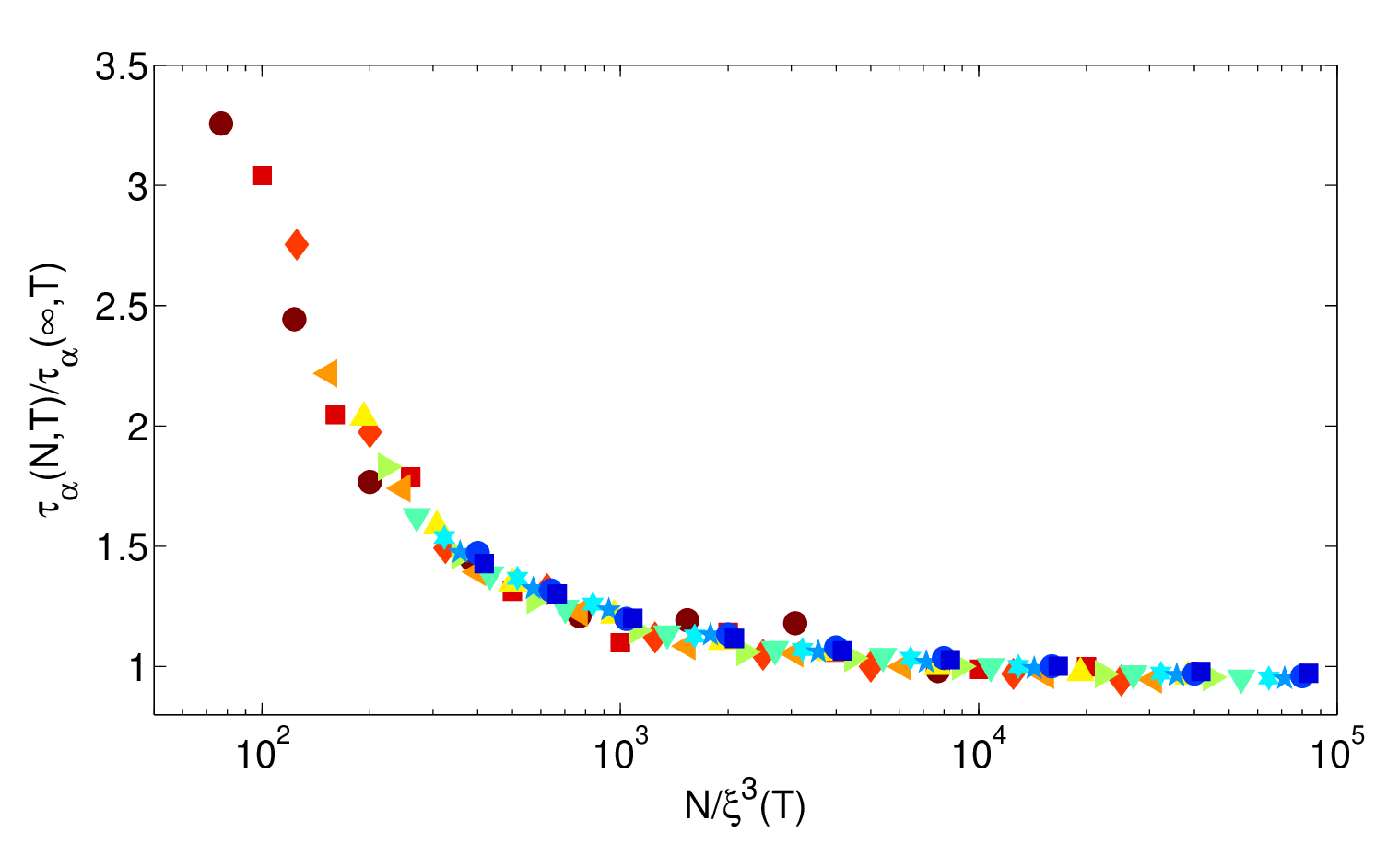}
	\caption{Upper panel: System size dependence of relaxation time	rescaled by the relaxation time of the largest system. The system size dependence becomes more enhanced at lower temperatures,	indicating the existence of a lengthscale.	Note that in these MD simulations the system sizes used span from $N=100$
to $N=20,000$. Lower panel: Collapse of the same data to obtain the lengthscale.}
		\label{tauScaling}
\end{figure}

The Hessian-based analysis
becomes less reliable at higher temperatures due to anharmonic effects. Thus for higher temperatures we employ yet another approach to determine the lengthscale $\xi$, extending its range of validity even further.
In Ref.~\cite{13BKP}, it was
argued that the point-to-set method  offers a computation of the lengthscale at high
temperatures. It was also found that finite size effects of the relaxation times $\tau$ at
different temperatures are perfectly correlated to the point-to-set lengthscale. This was tested using the lengthscale obtained in Ref.
~\cite{13BKP} for the Kob-Andersen model system. Since a finite size scaling
analysis of the relaxation time as in Ref.~\cite{12KP} is significantly easier
than calculating the point-to-set lengthscale, we employ the former to extract the static lengthscale at higher
temperatures as explained next.

In the upper panel of Fig.~\ref{tauScaling}, we show the system size
dependence of the relaxation time measured by molecular dynamics, rescaled by the value for the largest $N$ considered.
The scattered data themselves are a first indication that even at higher temperatures the
effect of the static lengthscale is substantial. In the lower panel we have
collapsed the data by appropriately choosing the lengthscale.

In both the minimal eigenvalue and the finite size $\tau$ scaling methods the lengthscales are defined up to a constant factor. For the latter method we choose this factor so that the lengthscale is unity at the highest temperature considered. The factor for the minimal eigenvalue method is chosen appropriately, considering that the results obtained from both methods should match at some
intermediate temperature range. In Fig.~\ref{fits} upper panel, we show the combined results of both methods. One can
clearly see that the static lengthscale varies by more than 500\%, which is quite remarkable.

\begin{figure}
	\vskip 0.3 cm
    \onefigure[scale = 0.34]{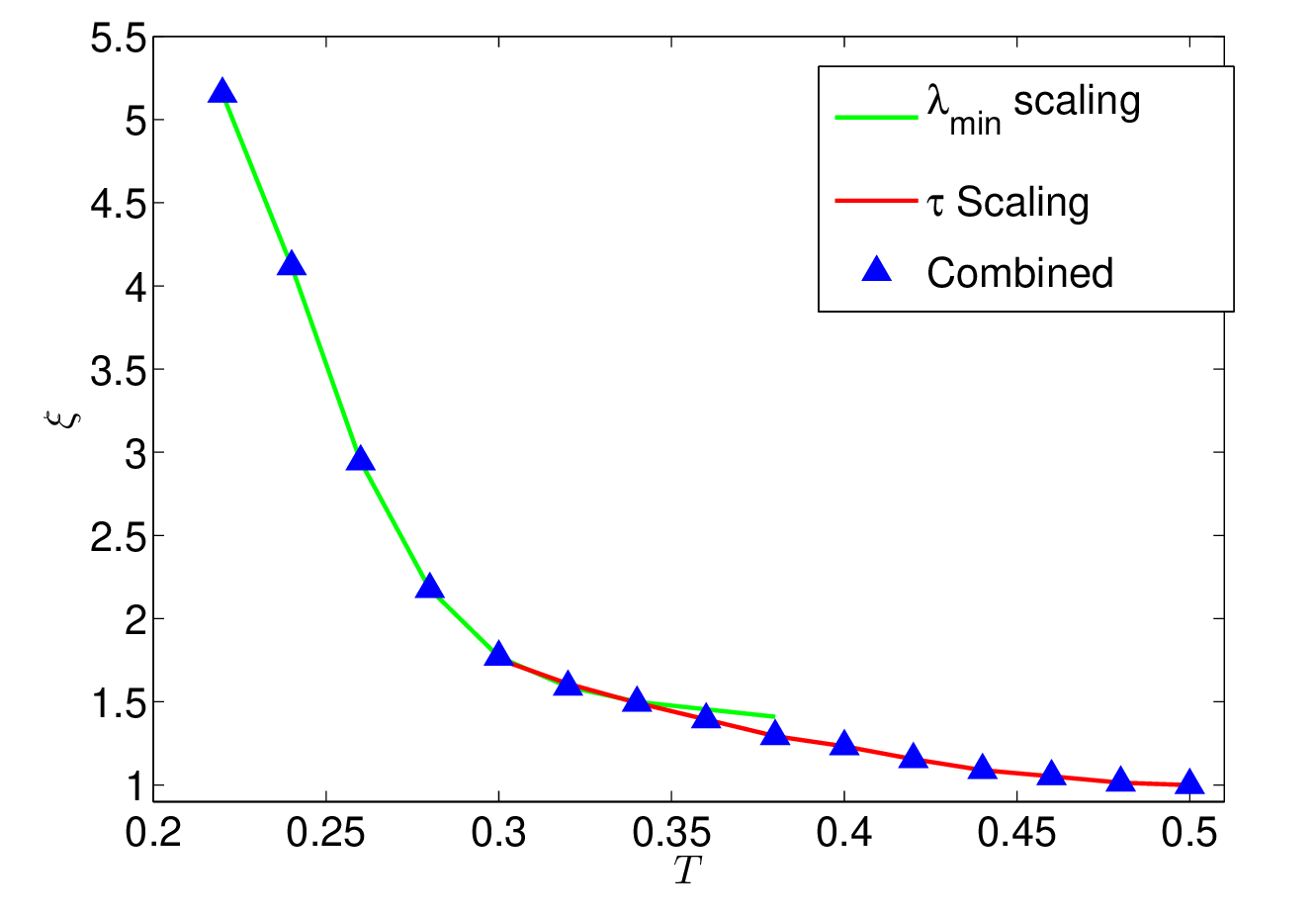}
	\onefigure[scale = 0.25]{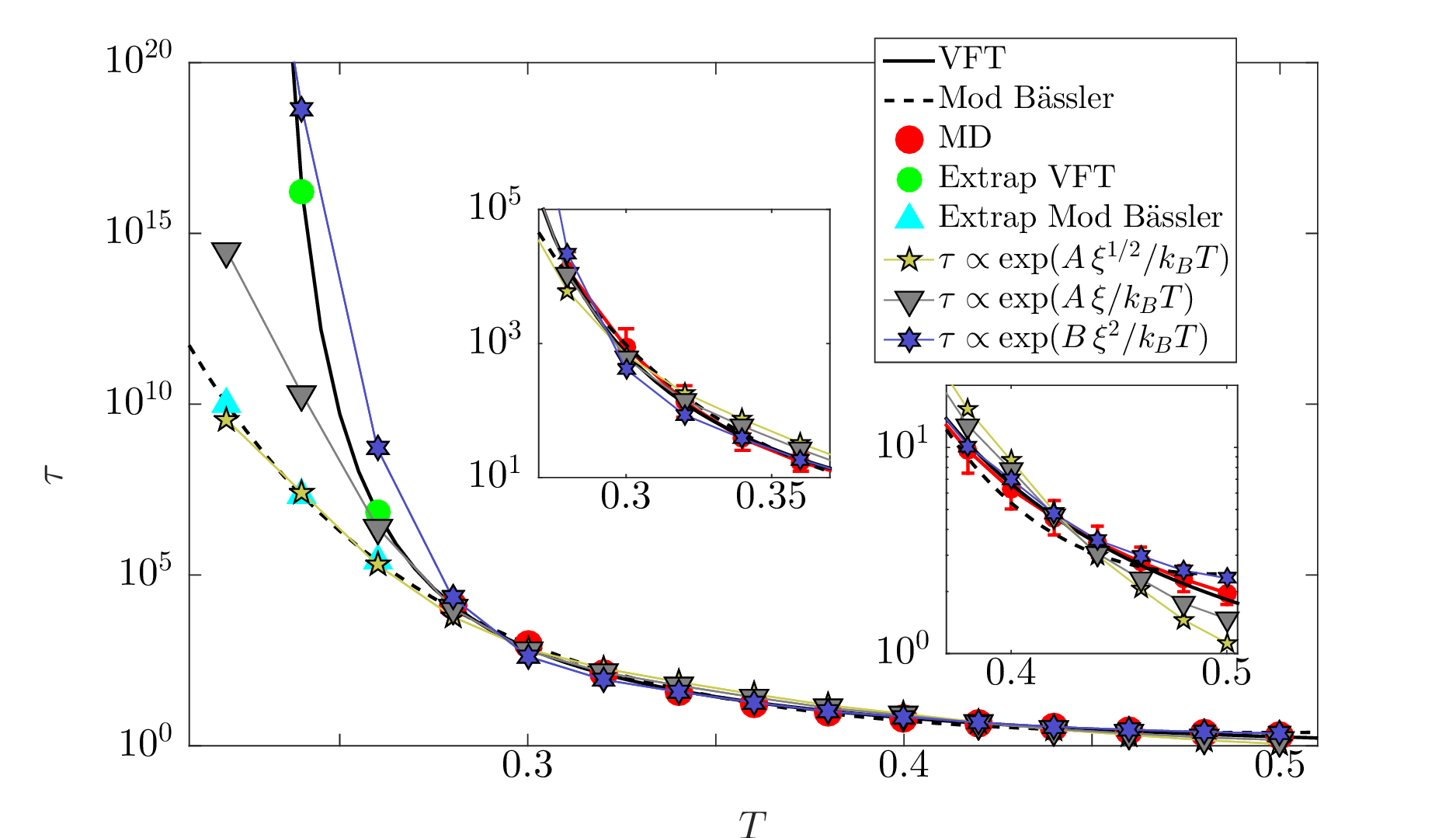}
	\caption{Upper panel: The lengthscale $\xi(T)$ for the whole temperature range. Lower
		Panel: Comparison of various fits and predictions of the relaxation times.
		In red dots we present the relaxation times measured via direct molecular
		dynamics simulations. The inset shows the VFT and B\"assler fits
		to our relaxation times measured data. The main figure shows the same and in
		addition the prediction of both these formulae for lower temperatures.
		Finally we also show a fit of the form $\tau = \tau_0 \exp [C\xi(T)/T]$  in a grey continuous
		line in addition to its predictions for $T=0.22$, $T= 0.24$ and $T=0.26$.}
		\label{fits}
\end{figure}

Presently we relate the results of the measured lengthscale $\xi(T)$
to the $\alpha$ relaxation time $\tau$ as measured from the intermediate
scattering function for temperatures $T\geq0.28$ (and to extrapolations for lower $T$).
Our values of $\tau(T)$ were measured by averaging over at least fifty relaxation times.
Traditionally the relaxation times are fitted to the Vogel-Fulcher-Tamman (VFT) formula:
$
%\begin{equation}
\tau = \tau_0 \exp\left(\frac{A}{T-T_0}\right) \ .
\label{VF}
%\end{equation}
$
We fitted the three parameters in this formula to our measured relaxation
times in the range of temperatures $T\in [0.28,0.50]$,
as shown in Fig.~\ref{fits} lower panel, continuous black line. While this formula fits well the
{\it measured} relaxation time,
it predicts a divergence of the relaxation time
at $T_0=0.23$. Since we succeed to equilibrate the system at $T=0.22$ we
propose that this formula is untenable for the present system, ruling out
the VFT fit.

Another proposed fit formula is the so called modified B\"assler proposition
~\cite{09ECG},
$
\tau = \tau_0 \exp\left[B \left(\frac{T_1}{T}-1\right)^2\right] \ ,
%\label{Bass}
$
which is again a three parameter fit.
Fitting again to our available relaxation times
we find the dashed line in Fig.~\ref{fits} lower panel. The small deviation of this fit at temperatures above $T_1  = 0.48$ is expected, and has been
attributed to the facilitated particle motion at those high temperatures
~\cite{09ECG}.
However since the relaxation time of $4.6\times 10^5$ expected by this fit at $T=0.26$ is within the reach of our simulation (spanning $10^7$ time units), and we were not able to equilibrate the system by molecular dynamics at that temperature (see upper panel of Fig.~\ref{eq_check}), we can rule out also this extrapolation.
Indeed, the authors in Ref.~\cite{09ECG} predict the existence of a lower
bound for the range of validity of this fit providing a reason for the inadequacy of this extrapolation.

Finally we refer to the fit formula proposed in Ref.~\cite{12KP}:
\begin{equation}
\label{our}
\tau = \tau_0 \exp\left[\frac{C\xi(T)}{T}\right] \ ,
\end{equation}
which is a 2-parameter fit since $\xi(T)$ is determined. Note that in previous publications it was unclear whether the length scale $\xi$ in the exponent in Eq.~\ref{our} should be raised to power $d$ or not. In Fig.~\ref{fits} we also show the predictions of the same formula with $\xi$ raised to power 1/2 and 2.
The former is very close to the B\"assler prediction, and the latter predicts relaxation times
 that are two long. The fit with 1/2 can again be dismissed since it predicts a relaxation time that
 could be reached by MD simulations where we know that equilibration was not achieved. The fit with 2 overestimates the relaxation time even at $T=0.28$ where MD data exists.  Assuming that the Eq.~\ref{our} is applicable, and determining the parameters from the measured
relaxation times, we obtain the curve shown in a continuous grey line in Fig.~\ref{fits} lower panel. While this is not
rigorously justified at this time, the reader should note
that at the lowest temperature one
predicts $\tau= 1.64 \times 10^{15}$, which if correct extends the
analysis presented here to a range of $15$ orders of magnitude in relaxation
times.

In conclusion, we have presented a calculation of the typical static lengthscale $\xi$ in a range of temperatures that spans the interval $T\in [0.22,0.50]$, with an increase in $\xi$ by a factor of about 5, being unprecedented at this point in time.
Whichever model relating this lengthscale
to the relaxation time $\tau$ one takes, this increase in $\xi$  is associated with an increase in $\tau$ by a factor between
$10^{12}$ and $10^{16}$, meaning that we are able to offer an analysis that is competitive with the best experiments.

\acknowledgments
The authors would like to thank Giulio Biroli, Giorgio Parisi and Francesco Zamponi  for useful
discussions regarding finding the lengthscale at lower temperatures.


\begin{thebibliography}{0}

\bibitem{BookDH}
  \Book{Dynamical heterogeneities in glasses, colloids and granular materials}
  \Editor{ Berthier L.,  Biroli G.,  Bouchaud J.-P.,  Cipelletti L. \and van Saarloos  W.}
  \Publ{Oxford University Press}
  \Year{2011}

\bibitem{95HH}
  \Name{Hurley M. M.  \and Harrowell P. }
  \REVIEW{Phys. Rev. E.}{52}{1995}{ 1694}.

\bibitem{99BDBG}
\Name{ Bennemann C.,  Donati C. , Baschnagel  J. \and  Glotzer S. C.}
\REVIEW{ Nature} {399}{1999}{246}.

\bibitem{02DFGP}
\Name{Donati C. ,  Franz S., Glotzer  S. C. \and Parisi G.}
\REVIEW{ J. Non-Cryst. Solids} {307-310}{2002}{215} .

\bibitem{00FP}
\Name{ Franz S. \and  Parisi G.}
\REVIEW{ J. Phys. Condens. Matter.} {12}{2000}{ 6335}.


\bibitem{05Chi4}
\Name{ Berthier L.,  Biroli G.,  Bouchaud J.-P.,  Cipilletti L.,  El Masri D.,  L'H\^ote D.,  Ladieu F. \and  Pierno M.}
\REVIEW{Science} {310}{2005} {1797} .

\bibitem{PREchi4}
\Name{  Toninelli C.,  Wyart M.,  Berthier L.,  Biroli G., \and  Bouchaud J.-P.}
\REVIEW{Phys. Rev. E} {71}{2005}{041505} .

\bibitem{09KDS}
\Name{ Karmakar S.,  Dasgupta C., \and  Sastry S.}
\REVIEW{ Proc. Nat. Acad. Sci. (USA)} {106}{2009}{3675} .

\bibitem{10KDS}
\Name{ Karmakar S.,  Dasgupta C., and  Sastry S.}
\REVIEW{ Phys. Rev. Lett. }{ 105}{2010}{015701} .

\bibitem{11BB}
\Name{ Berthier L. \and  Biroli G.}
\REVIEW{ Rev. Mod. Phys.} {83}{2011}{587 }.

\bibitem{14KDS}
\Name{ Karmakar S. ,  Dasgupta C. \and  Sastry S.}
\REVIEW{  Annu. Rev. Condens. Matter Phys.}  { 5}{2014}{255} .

\bibitem{01Donth}
\Name{ Donth E.}
\Book{The Glass Transition}
\Publ{Springer, Berlin}
\Year{ 2001}.

\bibitem{01DS}
\Name{ Debenedetti P. G. \and Stillinger  F. H.}
\REVIEW{ Nature} {410} {2001}{259 }.

\bibitem{04BB}
\Name{ Bouchaud J-P. \and  Biroli G.}
\REVIEW{ J.Chem.Phys.} { 121}{2004}{ 7347} .


\bibitem{MM}
\Name{ M\'ezard M. \and  Montanari A.}
\REVIEW{ J. Stat. Phys.} {124}{2006}{1317 } .

\bibitem{CGV}
\Name{Cavagna A. , Grigera  T. S., \and Verrocchio  P.}
\REVIEW{ J. Chem. Phys.} {136} {2012}{204502}.

\bibitem{08BBCGV}
\Name{ Biroli G.,  Bouchaud J.-P.,  Cavagna A.,  Grigera T.S. \and  Verrocchio P.}
\REVIEW{ Nature Phys.} { 4} {2008}{771}  .

\bibitem{SauTar10}
\Name{Sausset F.  \and  Tarjus G.}
\REVIEW{ Phys. Rev. Lett.} { 104} {2010}{ 065701}.

\bibitem{BeKo_PS}
\Name{ Berthier L. \and  Kob W.}
\REVIEW{ Phys. Rev. E }{ 85} {2012}{011102}.

\bibitem{HocRei12}
\Name{ Hocky G.M.,  Markland T.E. \and  Reichman D.R.}
\REVIEW{ Phys. Rev. Lett.} {108}  {2012}{ 225506}.

\bibitem{11KLP}
\Name{ Karmakar S., Lerner E. \and  Procaccia I.}
 \REVIEW{ Physica A} { 391} {2012}{ 1001}.

\bibitem{12KP}
\Name{ Karmakar S. \and Procaccia I. }
\REVIEW{ Phys. Rev. E} { 86} {2012}{ 061502}.

\bibitem{13BKP}
\Name{ Biroli G.,Karmakar  S. \and  Procaccia I.}
\REVIEW{ Phys. Rev. Lett.} { 111} {2013}{ 165701}.

\bibitem{02TWLB}
\Name{ Tanguy A.,  Wittmer J.P.,  Leonforte F. \and  Barrat J.-L.}
\REVIEW{ Phys.Rev. B} { 66} {2002}{ 174205}.

\bibitem{10Sok}
\Name{ Sokolov A.}
http://online.kitp.ucsb.edu/online/glasses-c10/sokolov/



\bibitem{09IPRS}
\Name{ Ilyin V.,Procaccia I. , Regev I. \and  Shokef Y.}
\REVIEW{ Phys. Rev. B} { 80}  {2009}{174201}.

\bibitem{11HKLP}
\Name{ Hentschel H.G.E., Karmakar S. , Lerner E.  \and  Procaccia I.}
\REVIEW{  Phys. Rev. E} { 83}{2011}{061101} .

\bibitem{10KLP}
\Name{ Karmakar S.,  Lerner E. \and  Procaccia I.}
\REVIEW{  Phys.Rev. E} { 82} {2010}{055103(R)}.


\bibitem{01GP}
\Name{ Grigera T.S. \and  Parisi G.}
 \REVIEW{Phys. Rev. E} { 63}{2001}{ 045102(R)}.

\bibitem{04BR}
\Name{ Brumer Y. \and  Reichman D.R.}
 \REVIEW{J. Phys. Chem. B} { 108} {2004}{ 6832}.

\bibitem{07FMV}
\Name{ Fern\'andez L.A.,Mart\'in-Mayor V.  \and  Verrocchio P.}
\REVIEW{Philosophical Magazine} { 87}{2007}{ 581} .

\bibitem{00YK}
\Name{ Yamamoto R. \and  Kob W.}
\REVIEW{ Phys. Rev. E} { 61} {2000}{ 5473}.

\bibitem{07PTVF}
\Name{ Press W.H.,  Teukolsky S.A., Vetterling  W.T., \and  Flannery B.P.}
\Book{ Numerical Recipes: The Art of Scientific Computing, 3rd Edition}
\Publ{Cambridge University Press, New York}
\Year{2007}

\bibitem{96GL}
\Name{ Golub  G.H.\and  van Loan C.F.}
\Book{Matrix Computations, 3rd Edition}
\Publ{Johns Hopkins University Press, City}
\Year{ 1996}


\bibitem{09ECG}
\Name{ Elmatad Y.S.,  Chandler D. \and  Garrahan J.P.}
\REVIEW{ J. Phys. Chem. B} { 113} {2009}{5563}; {\bf 114} {(2010)}{~17113}.

\bibitem{06MS}
\Name{ Montanari A. \and  Semerjian G.}
\REVIEW{ J. Stat Phys.} { 125} {2006}{ 23}.


\end{thebibliography}
\end{document}